\documentclass[preprint,aps,longbibliography,singlecolumn,superscriptaddress,amsmath,amssymb,verbatim]{revtex4}

\usepackage{graphicx}
\usepackage{lmodern}
\usepackage{epstopdf}
\usepackage{dcolumn}
\usepackage{bm}
\usepackage{subfigure}
\usepackage{amsmath} 
\usepackage{natbib}
\setcitestyle{super}
\usepackage[pdftex,colorlinks=true,citecolor=blue,linkcolor=blue,urlcolor=blue,bookmarks=true]{hyperref}

\def\@maketitle{%
	\newpage\spacing{1}\setlength{\parskip}{12pt}%
	{\Large\bfseries\noindent\sloppy \textsf{\@title} \par}%
	{\noindent\sloppy \@author}%
}

\usepackage{color}
\usepackage{textcomp}
\definecolor{red}{rgb}{1,0,0}

\definecolor{blue}{rgb}{0,0,1}

\definecolor{green}{rgb}{0,1,0}

\renewenvironment{abstract}{%
	\setlength{\parindent}{0in}%
	\setlength{\parskip}{0in}%
	\bfseries%
}{\par\vspace{-6pt}}

\begin{document}

	\title{Signature of a randomness-driven spin-liquid state in a frustrated magnet \vspace{-0.3 cm}}
	\author{J. Khatua}
	\affiliation{Department of Physics, Indian Institute of Technology Madras, Chennai 600036, India}
	\author{M. Gomil\v{s}ek}
	\affiliation{Jo\v{z}ef Stefan Institute, Jamova c. 39, 1000 Ljubljana, Slovenia}
	\affiliation{Faculty of Mathematics and Physics, University of Ljubljana, Jadranska u. 19, 1000 Ljubljana, Slovenia}
	\author{J. C. Orain}
	\affiliation{Paul Scherrer Institute, Bulk MUSR group, LMU 5232 Villigen PSI, Switzerland}
		\author{A. M. Strydom}
	\affiliation{Highly Correlated Matter Research Group, Department of Physics, University of Johannesburg, PO Box 524, Auckland Park 2006, South Africa}
	\affiliation{Max Planck Institute for Chemical Physics of Solids, 40 Nöthnitzerstr., Dresden D-01187, Germany.}
	\author{Z. Jagli\v{c}i\'{c} }
	\affiliation{Faculty of Civil and Geodetic Engineering, University of Ljubljana, 1000 Ljubljana, Slovenia}
	\affiliation{Institute of Mathematics, Physics and Mechanics, 1000 Ljubljana, Slovenia}
\author{C. V. Colin}
\affiliation{Institut Néel, Université Grenoble Alpes, CNRS, Grenoble, 38042, France}
\author{S. Petit}
\affiliation{LLB, CEA, CNRS, Université Paris-Saclay, CEA Saclay, 91191 Gif-sur-Yvette, France}
	\author{A. Ozarowski}
	
	\affiliation{National High Magnetic Field Laboratory, Florida State University, Tallahassee, Florida 32310, USA}
		\author{L. Mangin-Thro}
	\affiliation{Institut Laue-Langevin, 38042 Grenoble, France}
	\author{K. Sethupathi }
	\affiliation{Department of Physics, Indian Institute of Technology Madras, Chennai 600036, India}
	\affiliation{Quantum Centre for Diamond and Emergent Materials, Indian Institute of Technology Madras,
		Chennai 600036, India.}
	\author{ M.S. Ramachandra Rao}
	\affiliation{Quantum Centre for Diamond and Emergent Materials, Indian Institute of Technology Madras,
		Chennai 600036, India.}
	\affiliation{Department of Physics, Nano Functional Materials Technology Centre and Materials Science Research Centre,
		Indian Institute of Technology Madras, Chennai-600036, India}
	\author{A. Zorko}
	\affiliation{Jo\v{z}ef Stefan Institute, Jamova c. 39, 1000 Ljubljana, Slovenia}
	\affiliation{Faculty of Mathematics and Physics, University of Ljubljana, Jadranska u. 19, 1000 Ljubljana, Slovenia}
	\author{P. Khuntia}
	\affiliation{Department of Physics, Indian Institute of Technology Madras, Chennai 600036, India}
	\affiliation{Quantum Centre for Diamond and Emergent Materials, Indian Institute of Technology Madras,
		Chennai 600036, India.}
	\affiliation{Functional Oxide Research Group, Indian Institute of Technology Madras, Chennai 600036,
		India.}
	
	
		\maketitle
	\begin{abstract}
		Collective behaviour of electrons, frustration induced quantum fluctuations and entanglement in quantum materials underlie some of the emergent quantum phenomena with exotic quasi-particle excitations that are highly relevant for technological applications. Herein,  we present our thermodynamic and muon spin relaxation measurements, complemented by \textit{ab initio} density functional theory and exact diagonalization results,  
		on the recently synthesized frustrated  antiferromagnet Li$_{4}$CuTeO$_{6}$, in which Cu$^{2+}$ ions ($S$ = 1/2) constitute disordered spin chains and ladders along the crystallographic [101] direction with weak random inter-chain couplings. Our thermodynamic experiments detect neither long-range magnetic ordering  nor spin freezing down to 45 mK despite the presence of strong antiferromagnetic interaction between Cu$^{2+}$ moments  leading to  a large effective Curie-Weiss temperature of $–$154 K. Muon spin relaxation results are consistent with thermodynamic results. The temperature and magnetic field scaling of magnetization and specific heat reveal a data collapse pointing towards the presence of random-singlets within a disorder-driven correlated and  dynamic ground-state in this frustrated  antiferromagnet.
	\end{abstract}
\vspace{0.6 cm}
	Geometrically frustrated magnets with incompatible magnetic interactions between spins  are characterized by  strong quantum fluctuations and absence of trivial order parameter related to  broken symmetry.  They  are, therefore, extremely sensitive to perturbations such as quenched disorder, next-nearest-neighbor interactions, and magnetic anisotropy \cite{Balents2010,ANDERSON1973153}. Strong quantum fluctuations preclude symmetry breaking phase transitions and lead to novel magnetically disordered ground states such as quantum spin liquids (QSLs)
	\cite{Balents2010,PhysRevX.1.021002,Ramirez1999,KHUNTIA2019165435,PhysRevB.87.214417}.
QSLs are  characterized by quantum fluctuations, absence of long-range order  even at \textit{T}$\rightarrow{0}$, and exotic fractionalized excitations in contrast to spin-wave excitations found in conventional magnets \cite{Savary_2016,Castelnovo2008}. These properties of QSLs are fingerprints of strong quantum entanglement. Triangular-lattice compound 
$\kappa$-(ET)$_{2}$Cu$_{2}$(CN)$_{3}$ \cite{RevModPhys.89.025003} and 1-TaS$_{2}$ \cite{Klanjek2017},
kagome-lattice compound ZnCu$_{3}$(OH)$_{6}$Cl$_{2}$ \cite{Khuntia2020}, and hyperkagome-lattice compound PbCuTe$_{2}$O$_{6}$ \cite{PhysRevLett.116.107203,Chillal2020} 
are some of the  most investigated candidate spin-liquid materials with unusual ground-state properties. 
Recently, enormous efforts have been devoted towards experimental realizations of QSLs on the honeycomb-lattice ever since Alexei Kitaev proposed an exactly solvable highly anisotropic Kitaev model for $S$ = 1/2 spins on this  two dimensional spin-lattice \cite{KITAEV20062}. In sharp contrast to an isotropic nearest-neighbor exchange model  on the honeycomb-lattice, which exhibits long-range order \cite{Fouet2001}, the ground-state of the Kitaev model is a QSL with fractionalized Majorana fermions and gauge flux excitations \cite{Do2017,Takagi2019,PhysRevB.84.024406,Jansa2018}.  The realization of this exotic state of matter is interesting not only from a fundamental physics point of view but also holds immense promise for potential applications in robust quantum computing technology \cite{RevModPhys.80.1083}.\\ Despite tremendous experimental efforts, the
observed signatures of QSL are far away from those of theoretically proposed QSLs states for a clean periodic system \cite{ANDERSON1973153,PhysRevLett.59.2095,KITAEV20062}. However, it has been suggested that some  highly frustrated magnets show quantum disordered ground-state due to  randomness that might play an important role in destabilizing conventional N\'eel order at low $T$. There are several origins of randomness in magnetic insulators and the interplay between quantum fluctuation and randomness may lead to novel quantum ground-states. The triangular-lattice YbMgGaO$_{4}$ with anti-site mixing between non-magnetic Mg$^{2+}$ and Ga$^{3+}$ ions \cite{PhysRevX.8.031028,PhysRevLett.119.157201} and  square lattice Sr$_{2}$CuTe$_{1-x}$W$_{x}$O$_{6}$ with a few percent of W on the Te site are  some of examples of frustrated magnets with a randomness-induced  quantum disordered ground state \cite{PhysRevB.102.054443,PhysRevX.8.041040,PhysRevLett.126.037201}.   Interestingly, while most of the reported Kitaev magnets including 5\textit{d} iridates  A$_2$IrO$_{3}$ (where A = Na, Li) and ruthenate $\alpha$-RuCl$_{3}$ with 4$d$ electrons show long-range order at sufficiently low-temperature,  the iridate  H$_{3}$LiIr$_{2}$O$_{6}$ and ruthenate $\alpha$-Ru$_{1-x}$Ir$_{x}$Cl$_{3}$, which are  derived from a parent ordered magnet by substitution of non-magnetic ions, do not exhibit long-range magnetic order \cite{Kitagawa2018,PhysRevB.98.014407,PhysRevB.102.094407,PhysRevLett.124.047204}.  Furthermore, some honeycomb-lattice  magnets, like  A$_{3}$LiIr$_{2}$O$_{6}$ (A = H, D, Ag
etc.), show scaling behavior in the field dependence of
the specific heat \cite{Kitagawa2018,PhysRevLett.123.237203,Kimchi2018}, which is ascribed to the existence of
disorder, making these quantum materials interesting for
studies of the effect of disorder on a QSL state \cite{Kimchi2018}.  A similar scenario is  observed in three dimensional spin-lattice where quantum fluctuation is less pronounced. For instance, the oxynitride Lu$_{2}$Mo$_{2}$O$_{5}$N$_{2}$ with $S$ = 1/2 spins on a pyrochlore lattice, which is derived from the spin-glass material Lu$_{2}$Mo$_{2}$O$_{7}$ via random substitution of O$^{2-}$ and N$^{3-}$ anions, shows a dynamic ground state \cite{PhysRevLett.123.087201,PhysRevLett.113.117201}. \\
Disorder in quantum magnets is usually unavoidable, yet very often it can act as a new prism in revealing many interesting quantum phenomena of the host material \cite{Gomilsek2019,RevModPhys.81.45,Yamaguchi2017,PhysRevB.96.174411}. Frustrated magnets with quenched disorder in the form of material defects or a broad distribution of  exchange interaction strengths can exhibit a randomness induced spin-liquid state. Understanding the impact of quenched disorder on the quantum fluctuations that drive quantum disordered states in antiferromagnets is an interesting area of study \cite{PhysRevLett.48.344,PhysRevX.8.031028,PhysRevB.92.180411,PhysRevB.93.140408}. In the  so-called random-singlet phase,  singlets form due to a random distribution of antiferromagnetic exchange interaction between spin-1/2 defect sites, leading to a power-law distribution of exchange energies and density of states, which lead to unconventional scaling behavior in magnetic susceptibility and specific heat at low-temperatures \cite{PhysRevX.8.031028,Kimchi2018,doi:10.7566/JPSJ.86.044704}.
The signature of such novel random singlet state was initially observed in one-dimensional spin-chain organic compounds due to presence of random antiferromagnetic Heisenberg exchange interaction \cite{PhysRevLett.45.1303,PhysRevB.12.356}. Furthermore, there is also experimental evidence of random singlet state in inorganic spin-chain compound BaCu$_{2}$(Si$_{1-x}$Ge$_{x}$)$_{2}$O$_{7}$ \cite{PhysRevB.99.035116} wherein the random substitution of Ge in place of Si introduce the bond disorder in antiferromagnetic long-range ordered magnet BaCu$_{2}$Si$_{2}$O$_{7}$ \cite{PhysRevB.88.054422}. Theoretically, it has been proposed that such exchange randomness in spin-chain compound is one of the key ingredients for quantum information processor which deals with properties of quantum entanglement in quantum materials \cite{PhysRevLett.43.1434,PhysRevLett.106.040505,PhysRevB.94.174442,PhysRevB.97.104424}.  In recent years, higher dimensional quantum materials with intrinsic disorder either from spin vacancies, anti-site disorder or non-magnetic impurities and dislocations or grain boundaries have drawn enormous attention to realize novel low-temperature phases including random singlet that is expected to be unique in disordered quantum many-body systems \cite{PhysRevLett.123.087201,PhysRevLett.113.117201,PhysRevX.8.041040}.
 In frustrated magnets with quenched disorder, the majority of  intrinsic $S$ = 1/2 spins may  constitute a dynamic liquid-like state while the remaining  $S$ = 1/2  moments act as defects and  form a random network with exotic low-lying excitations that destabilize a glassy state and instead lead to a dynamic low-energy ground-state \cite{Kimchi2018}.
Theoretically, it has been proposed that randomness in the frustrated lattice can induce such  spin-liquid
like states in a wide region of parameter space \cite{Watanabe2014}. 
Experimentally, this phenomenon is much less explored due to a lack of suitable frustrated magnets, which leads one to look for candidate materials wherein the degree of randomness can be tuned in a controlled manner. In this context, 3$d$ transition-metal frustrated S = 1/2 spin lattices are promising potential hosts of a randomness driven quantum spin-liquid ground state.
\begin{figure*}[h!]
	\centering
	\includegraphics[width=\textwidth]{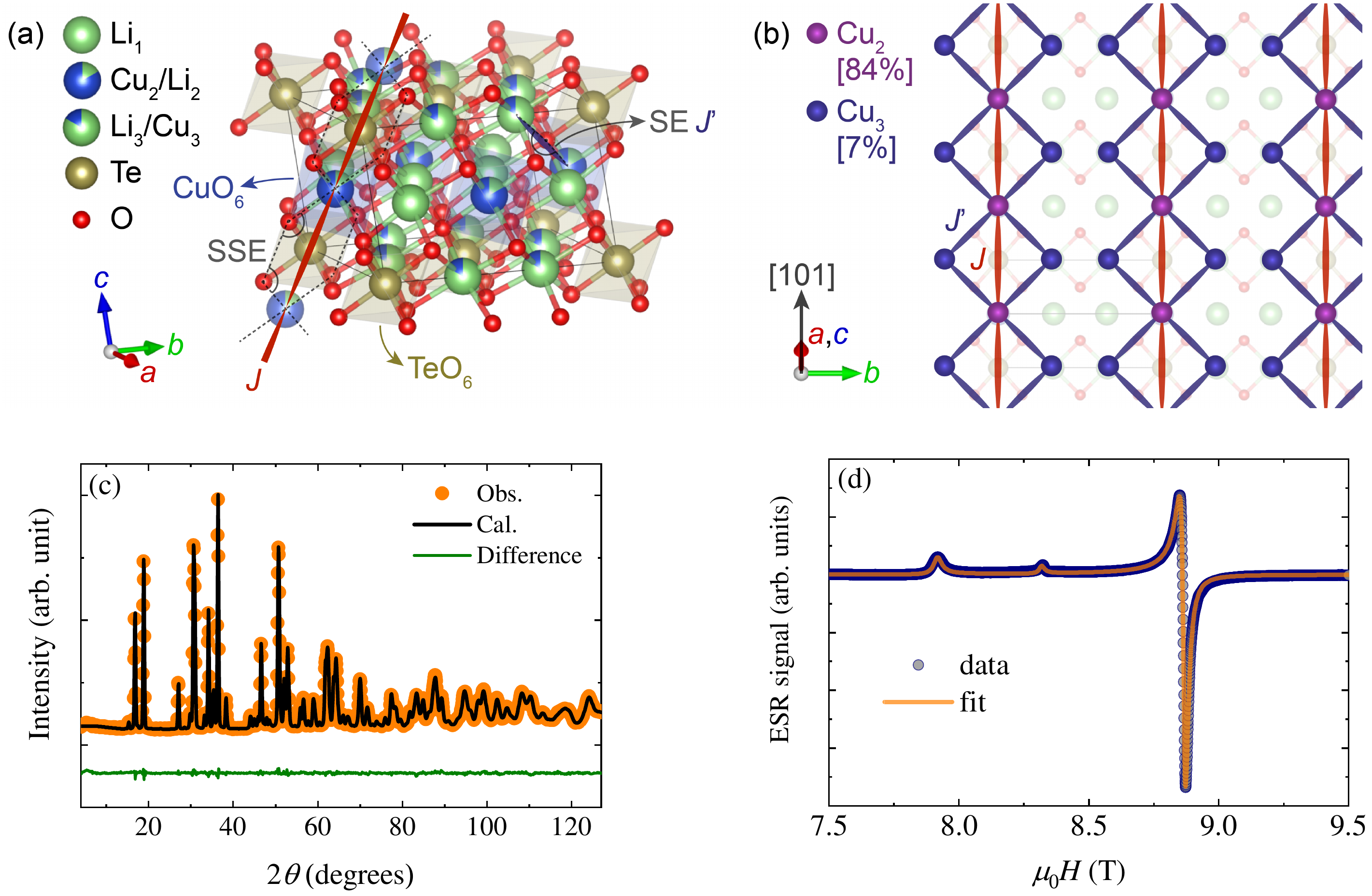}
	\caption{\textbf{Crystal structure and high temperature spin model of Li$_{4}$CuTeO$_{6}$}. (a) Visualization of one unit cell of LCTO where edge-sharing TeO$_6$ and CuO$_6$ octahedra connect Cu$^{2+}$ ions on $2d$ (Cu$_2$) sites with an exchange $J$ through a Cu--O${\cdots}$O--Cu super-superexchange (SSE) bridge around Te$^{6+}$. Additionally, corner-sharing CuO$_6$ octahedra connect Cu$^{2+}$ ions on $2d$ (Cu$_2$) and $4g$ (Cu$_3$) sites with an exchange $J'$ through a nearly-linear Cu--O--Cu superexchange (SE) bridge. (b) Resulting spin model of randomly-depleted 1D spin chains of Cu$_2$ sites with antiferromagnetic exchange $J$ running along the [101] direction, to which randomly-occupied Cu$_3$ sites couple via an antiferromagnetic exchange $J'$ and introduce strong frustration. (c)
		Rietveld refinement of  neutron  diffraction data of LCTO at room temperature for the incident wavelength 1.28 {\AA}. The solid circles represent  the observed intensity (Obs.) whereas the black solid line is the calculated intensity (Cal.). The residual signal is denoted by a green solid line  and is shifted vertically for clarity. (d) The ESR spectrum of LCTO measured at 250 K and 256.3 GHz (points). The fit (solid line) includes two components with uniaxial symmetry and a Lorentzian line shape.}{\label{Li$_{4}$CuTeO$_{6}$1}}
\end{figure*}
\\
Herein, we report crystal structure, neutron diffraction, magnetic properties, specific heat, and muon spectroscopy, as well as complementary density functional theory (DFT) and exact diagonalization (ED) results of the antiferromagnet Li$_{4}$CuTeO$_{6}$ (henceforth LCTO), where Cu$^{2+}$ ($S$ = 1/2) ions are arranged on a  frustrated spin-lattice with  unavoidable anti-site disorder between Li$^+$ and Cu$^{2+}$. In LCTO, the  majority of Cu$^{2+}$ ions (84 \text{\%}) at the 2$d$ crystallographic site form random-length strongly antiferromagnetic spin chains  while a minority of Cu$^{2+}$ ions ($\sim$ 7 \text{\%}) at defect 4$g$ sites strongly antiferromagnetically couple to these spin chains, leading to significant frustration. The  large and negative value of Curie-Weiss temperature reflects the presence of strong antiferromagnetic interaction between Cu$^{2+}$ spins.  Furthermore, we find that weaker inter-chain interactions effectively couple the chains in a random, disordered network at low $T$. As a consequence, this compound  neither undergoes a phase transition to long-range
magnetic order nor spin freezing down to  at least 45 mK. Furthermore, specific heat and magnetization results reveal data collapse behavior, which suggests the presence of a random-singlet state \cite{Kimchi2018}. Muon spin relaxation measurements corroborate a dynamic ground-state in this frustrated magnet. This is attributed to the presence of subdominant inter-chain interactions that couple the chains in a disordered network. Our results thus establish that  LCTO hosts a randomness induced spin-liquid-like state in a frustrated magnet.\\
\section*{Results}
\subsection*{Rietveld refinement  and crystal structure  }
The room-temperature powder X-ray diffraction data were taken to confirm the phase purity of the polycrystalline samples of LCTO. The Rietveld refinement of the X-ray diffraction data was performed using GSAS software \cite{Toby:hw0089}, which confirms that LCTO crystallizes in a monoclicnic structure with the space group $C2/m$ (No. 12) \cite{Kumar2012}. To deepen the analysis of the crystalline structure and in particular to determine the occupation of the mixed sites by the light elements a neutron diffractogram was measured at 300 K at an incident wavelength 1.28 \text{\AA}. Fig.~\ref{Li$_{4}$CuTeO$_{6}$1} (c) depicts the Rietveld refinement of neutron diffraction data performed using fullprof which  yields  lattice parameters  $a$ = 5.2752(5) {\AA}, $b$ = 8.8163(8) {\AA}, $c$ = 5.2457(5) {\AA} and $\alpha $ = $\gamma$ = 90$^{\circ}$,  $\beta$ = 113.172(8)$^{\circ}$, in agreement with powder X-ray diffraction data (see SI table I) and previously reported values \cite{Kumar2012}.
A unit cell of LCTO is shown in Fig.~\ref{Li$_{4}$CuTeO$_{6}$1} (a). We found that  among three crystallographic sites of Li, the sites 2\textit{d} and  4\textit{g} are partially shared with Cu$_{2}$ and Cu$_{3}$ sites, respectively, with Cu$_2$ site occupancy $p_{2}$ = 84 $\text{\%}$ and Cu$_3$ site occupancy $p_3$ = 7 $ \text{\%}$. This is likely due to the similar ionic radii of Li$^{+}$ and Cu$^{2+}$ ions.  
Cu$^{2+}$ ions  form distorted CuO$_{6}$ octahedra with nearest-neighbor oxygen ions with Cu$_2$ and Cu$_3$ sites connecting to each other directly by corner- and edge-sharing, and indirectly via corner- and edge-sharing TeO$_6$ octahedra in an \textit{a priori} 3D lattice. However, a critical look at the crystal structure reveals only two plausibly strong exchange pathways: (i) a nearly-linear Cu--O--Cu superexchange (SE) bridge between Cu$_2$ and Cu$_3$ sites with bond angle 173.3$^\circ$ and bond distance 4.12 \text{\AA}, which is expected to yield a strong antiferromagnetic exchange $J'$ by Goodenough-Kanamori rules~\cite{goodenough1976magnetism,rocquefelte2012theoretical}, and (ii) a double Cu--O${\cdots}$O--Cu super-superexchange (SSE) bridge between two Cu$_2$ sites going around a Te$^{6+}$ ion with a large symmetric Cu--O${\cdots}$O bond angle of 138.6$^\circ$ and a small O${\cdots}$O distance 2.87 \text{\AA} (well below the van der Waals radii sum $\sim$ 3.04 \text{\AA}), which is expected to yield a strong antiferromagnetic exchange $J$~\cite{rao2016tellurium,whangbo2003spin,whangbo2005spin,koo2008analysis,bertani2010halogen} [Fig. 1(a)]. The resulting 1D spin model is shown in Fig. 1(b) and consists of random-length linear spin chains of Cu$_2$ (occupation probability $p_2$) with exchange $J$ running along the [101] direction, with randomly-occupied Cu$_3$ defect sites (occupation probability $p_3$) displaced along $\pm b$ from each main-chain bond $J$ introducing strong frustration by coupling to both neighboring Cu$_2$ main-chain sites via a $J'$ exchange. All other SE and SSE pathways, which, if present, introduce 3D inter-chain couplings in general, are expected to have much smaller strength as they either proceed via very nearly 90$^\circ$ Cu--O--Cu and Cu--O${\cdots}$O angles, or involve O${\cdots}$O distances well above the O$^{2-}$ van der Waals radii sum.
\subsection*{\textit{Ab initio} and exact diagonalization calculations} The expected interaction between Cu$^{2+}$ moments are fully born out by \textit{ab initio} DFT exchange-coupling calculations on a $2 \times 1 \times 2$ supercell of LCTO using the local (spin) density approximation (LSDA) in an LSDA+$U$ scheme with an effective Hubbard $U_\mathrm{eff}$ = 9 meV  (see SI for details). By first assuming full occupancy of Cu$_2$ sites and zero occupancy of Cu$_3$ sites the main-chain Cu$_2$--Cu$_2$ antiferromagnetic exchange strength $J_\mathrm{DFT}$ = 177 K was extracted, and then by introducing a single occupied Cu$_3$ site the frustrating Cu$_2$--Cu$_3$ antiferromagnetic exchange strength $J'_\mathrm{DFT}$ = 824 K was obtained. As expected, all other exchange interactions, which introduce 3D couplings between the chains, were found to have strengths below $\sim$ 6 K, making them difficult to resolve by DFT. For $T \gg $ 6 K the frustrated 1D random-occupancy model of Fig. 1(b) is thus expected to hold, while at low $T$ further frustrating inter-chain exchanges might be expected. Temperature dependent thermodynamic quantities of the high-$T$ random model obtained from DFT were calculated using ED by probability weighing ED results for different random-length chains of Cu$_2$ spins with a different distribution of occupied Cu$_3$ sites around them (see SI for details). This way, ED predictions for magnetic susceptibility $\chi_\mathrm{ED}$, magnetic specific heat $C_\mathrm{mag,ED}$ and magnetic entropy $S_\mathrm{mag,ED}$ as a function of $T$, applied magnetic field $\mu_0 H$, $g$ factor, and $J$ and $J'$ exchange strengths at any given occupation probabilities $p_2$ and $p_3$ could be calculated to high accuracy.  
\subsection*{Magnetic susceptibility}
Fig.~\ref{Li$_{4}$CuTeO$_{6}$2} (a) presents the experimental temperature dependence of the magnetic susceptibility ($\chi(T)$) in  an applied magnetic field $\mu_{0}H$ = 1 T in the temperature range 1.9 K $\leq$ $T$ $\leq$ 300 K. The magnetic susceptibility data do not indicate the presence of a  phase transition down to 1.9 K. In order to estimate dominant magnetic interactions,  the high-temperature 1/$\chi(T)$ data were fitted  with a Curie-Weiss law (right $y$- axis; Fig.~\ref{Li$_{4}$CuTeO$_{6}$2} (a))
\begin{equation}\label{eq:1}
	\chi= \frac{C}{\textit{T}-\theta_{\rm CW}}.  
\end{equation} In Eq.~(\ref{eq:1}),
$\theta_{\rm CW}$ and
 \textit{C} are the Curie-Weiss temperature and the Curie constant, respectively. The Curie-Weiss fitting in the temperature range 180 K $\leq$ \textit{T} $\leq$ 300 K yields  $\theta_{\rm CW}$ = $-$154 K and \textit{C} = 0.39 cm$^{3}$ K/mol. The estimated effective magnetic moment $\mu_{\rm eff}$ = $\sqrt{8C}$ = 1.76 $\mu_{\rm B}$ is close to the expected value for a Cu$^{2+}$ ion with $S$ = 1/2 spin. The large negative $\theta_{\rm CW}$ confirms the presence of dominant antiferromagnetic interactions between Cu$^{2+}$ spins. 
\begin{figure*}[h!]
	\centering
	\includegraphics[width=\textwidth]{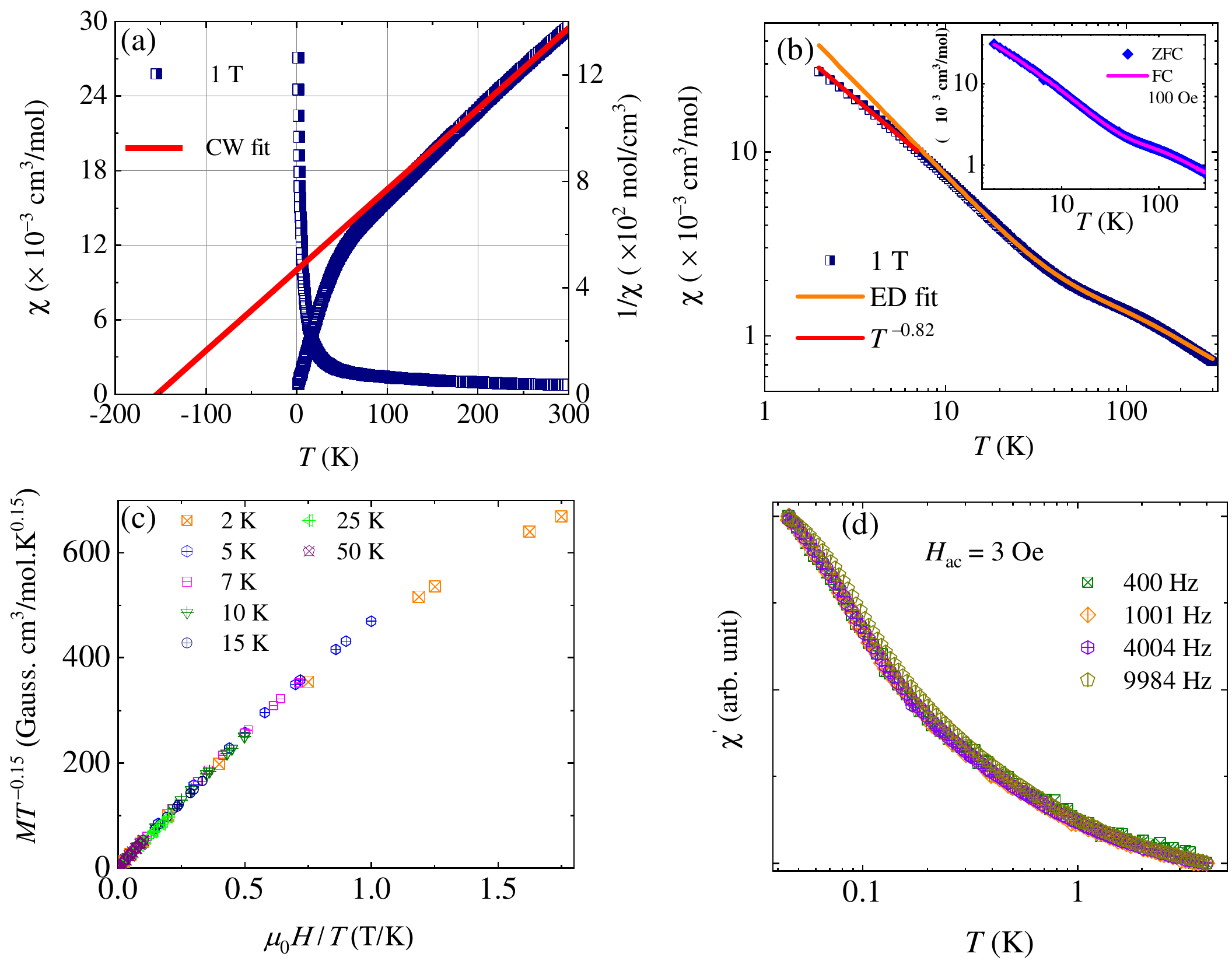}
	\caption{\textbf{Magnetization results of Li$_{4}$CuTeO$_{6}$}. (a) The temperature dependence of the magnetic
		susceptibility (left y-axis) and inverse magnetic susceptibility (right y-axis) in a magnetic field $\mu_{\rm 0}H$ = 1 T. The solid line represents the Curie-Weiss fit. (b) The temperature dependence of magnetic susceptibility at 1 T with a high-$T$ fit using ED results, and a low-$T$ fit using a power-law.  The inset shows the zero-field cooled (ZFC) and filed cooled (FC) susceptibility in a magnetic field of 100 Oe. (c) Data collapse for $MT^{-0.15}$ as a function of $\mu_{0}H/T$. (d) The temperature dependence of the real part of ac susceptibility in different frequencies down to 45 mK.}{\label{Li$_{4}$CuTeO$_{6}$2}}
\end{figure*} 
To obtain a more quantitative estimate of the $J$ and $J'$ exchange strengths, the $\chi(T)$ data were next fitted with the results of ED calculations on the random spin model of Fig. 1(b). Fixing the ratio of experimental $J/J'$ to be the same as that found in DFT, $J_\mathrm{DFT}/J'_\mathrm{DFT}$, which is expected to be a more robust DFT prediction than raw exchange strengths, and fixing the main-chain Cu$_2$ site occupancy to the experimental value $p_2$ = 84 \text{\%}, a good fit at $T > $ 20 K is obtained [Fig. 2(b)] with only 3 free, sensible parameters: $g = 2.28$, $J/J_\mathrm{DFT} = J'/J'_\mathrm{DFT} = 1.38$, and $p_3 = $ 20 \text{\%}. We note that while fit parameters $g$ and $J/J_\mathrm{DFT} = J'/J'_\mathrm{DFT}$ are quite robust to perturbations like magnetic anisotropies or unwanted extrinsic contributions due to any impurities, the fitted value of $p_3$ is substantially less so since the experimental data is limited to $T \ll J'$. Thus, the experimental value of Cu$_{3}$ occupancy from neutron diffraction should be taken as definitive. The final estimates for the dominant exchanges are thus $J$ = 244 K and $J'$ = 1140 K, with the discrepancy from DFT likely arising from the choice of Hubbard $U_\mathrm{eff}$, which acts to rescale all exchanges. Importantly, for $T < $ 20 K the high-$T$ model of Fig. 1(b) starts to systematically deviate from experiment, indicating the presence of further, most likely inter-chain, exchanges $J_\mathrm{inter} \lesssim $ 10 K in strength acting between the random spin-chain fragments of the high-$T$ model and dominating the physical response of LCTO at low $T < 20 \  \rm K \ll$ $J,J '$. In this regime most spin degrees of freedom of spin-chain fragments are expected to freeze out, leaving only their ground state degrees of freedom to couple via $J_\mathrm{inter}$ exchanges in an emergent $\geq 2$ dimensional (likely 3D) random lattice.
Remarkably, though LCTO  exhibits such quenched disorder, ultimately arising from partial Cu$^{2+}$ site occupancy,  that could lead to a spin-glass state at low $T$, the absence of glass-like freezing behavior is witnessed by the absence of any separation between zero-field-cooled and field-cooled susceptibility measurements in 100 Oe (inset of Fig.~\ref{Li$_{4}$CuTeO$_{6}$2} (b)). Furthermore, to confirm the absence of spin-freezing, we have performed ac susceptibility measurement down to 45 mK at different frequencies. It is observed that there is no peak or frequency dependence of real part of the ac susceptibility (Fig.~\ref{Li$_{4}$CuTeO$_{6}$2} (d) and SI Fig.~3) in the temperature range 45 mK $\leq$ $T$ $\leq$ 50 K, which strongly rules out the presence of spin glass transition in Li$_{4}$CuTeO$_{6}$ \cite{PhysRevB.62.6521}. 
Quenched Li/Cu anti-site disorder in the host lattice could, alternatively, also lead to unconventional scaling behavior of thermodynamic observables. Such a scenario has been proposed in several frustrated quantum materials \cite{PhysRevX.8.041040,PhysRevX.8.031028,Kimchi2018,PhysRevLett.125.117206}.
\\
As shown in Fig.~\ref{Li$_{4}$CuTeO$_{6}$2} (c),
a data collapse of $MT^{-0.15}$ as a function of $\mu_{\rm 0}H/T$ is found in LCTO in a broad temperature range between 2 and 50 K, which could be a sign of the presence of  random-singlets \cite{PhysRevLett.125.117206}. 
\begin{figure*}[h]
	\centering
	\includegraphics[width=\textwidth]{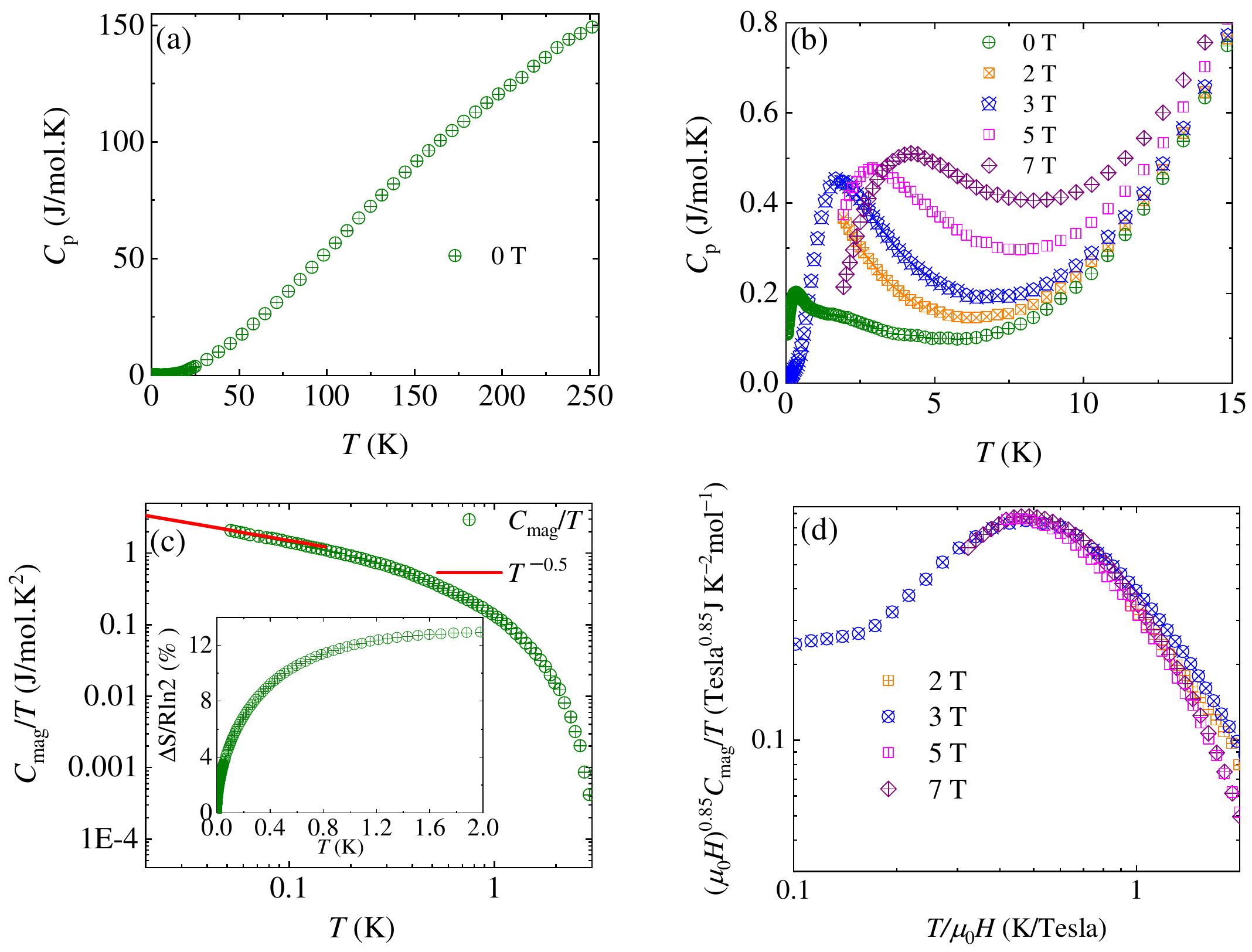}
	\caption{\textbf{Specific heat measurements reveal a randomness induced spin-liquid state in Li$_{4}$CuTeO$_{6}$ at low $T$}. (a) The temperature dependence of the specific heat ($C_{\rm p}$) in zero magnetic field in a broad temperature range 0.052 K $\leq$ \textit{T}  $\leq$ 250 K.   (b) The temperature-dependent specific heat in different magnetic fields upto 7 T. (c) The temperature dependence of magnetic specific heat divided by temperature in zero-field in a log-log scale where the red line is the power-law fit. Inset shows the temperature dependence of associated entropy change.  (d) The scaling behavior of the specific heat with $C_{\rm mag}(\mu_{0}H)^{0.85}/T$  as a function of the scaled temperature \textit{T}/$\mu_{0}H$ for several fields. }{\label{Li$_{4}$CuTeO$_{6}$3}}
\end{figure*} 
Furthermore, theoretically, it has been suggested  that in the absence of the spin-orbit interaction, low-temperature magnetic susceptibility data of random-singlet states show a power-law $\chi \propto T^{-\gamma}$ behavior with $\gamma$ $< 1$ \cite{PhysRevB.102.054443}. 
The susceptibility data were fitted with a power-law  $\chi \propto T^{-\gamma}$ in the low-temperature region, which yielded $\gamma$ = 0.82 $ \pm$ 0.01 (Fig.~\ref{Li$_{4}$CuTeO$_{6}$2} (b)).
A similar power-law fit was also observed in the quasi-two dimensional magnet Sr$_{2}$CuTe$_{x}$W$_{1-x}$O$_{6}$, which is a promising candidate for realizing a  random-singlet state on a  square lattice \cite{PhysRevX.8.041040}.  
A power-law fit of low-temperature magnetic susceptibility thus suggests the realization of a random-singlet state attributed to  the presence of isolated magnetic moments in the host lattice \cite{PhysRevX.8.041040,Kimchi2018}.
The data collapse is a remarkable feature found also in other QSL candidates with quenched disorder such as the honeycomb-lattice  H$_{3}$LiIr$_{2}$O$_{6}$ \cite{Kitagawa2018} and the triangular-lattice Y$_{2}$CuTiO$_{6}$ \cite{PhysRevLett.125.117206}. The data collapse behavior in the low-temperature region suggests the presence of random-singlets in LCTO induced by quenched disorder as  predicted theoretically \cite{Kimchi2018}. 
The powder ESR spectrum consists of two components, both featuring  uniaxial symmetry, in agreement with the crystal symmetry. A fit of the spectrum with a powder-averaged Lorentzian shape (see Fig.~\ref{Li$_{4}$CuTeO$_{6}$1} (d)) yields principal $g$ factors $g_{x,y}^{1}$ = 2.06 and $g_{z}^{1}$ = 2.31 for the component with larger intensity and $g_{x,y}^{2}$ = 2.06 and $g_{z}^{2}$ = 2.20 for the component with smaller intensity. The ratio of the intensities is 85:15, which is close to the ratio $p_2 : 2p_3$ of the occupied Cu$_2$ and Cu$_3$ sites as confirmed from the structural refinement.
\subsection*{Specific heat} 
Fig.~\ref{Li$_{4}$CuTeO$_{6}$3} (a) depicts the temperature dependence of  specific heat in zero magnetic field. The absence of any  anomaly in  the entire temperature range of investigation rules out any phase transition down to 52 mK, which is much below both the Curie-Weiss temperature and the dominant $J$ and $J'$ exchange interactions and thus indicates the presence of strong spin frustration that promotes a dynamic ground-state in LCTO.   
The specific heat results in several magnetic fields up to 7 T are shown in Fig.~\ref{Li$_{4}$CuTeO$_{6}$3} (b). It is observed that upon increasing the magnetic field the broad maximum shifts toward higher temperatures.
Though similar to a Schottky contribution, which could arise from quasi-free spins, it, in fact, describes intrinsic spin physics due to both spins that appear isolated in the high-$T$ spin model [Fig. 1(b)], and due to finite-length spin-chain fragments with an odd total number of spins, as the total spin of those cannot be lower than $1/2$ by the rules for quantum addition of spin. In both cases, an applied field Zeeman splits the spin-$1/2$ ground-state introducing a local-field-dependent energy gap that results in a specific heat peak whose position is proportional to the applied field. However, all spins, both those in finite-length spin-chain fragments, as well as nominally isolated spins, in fact couple together to form a frustrated 3D random spin-lattice via weaker $J_\mathrm{inter}$ inter-chain exchanges, together shaping the intriguing, liquid-like low-$T$ behaviour of LCTO at $T \lesssim J_\mathrm{inter}$. \\
\begin{figure*}
	\centering
	\includegraphics[width= \textwidth]{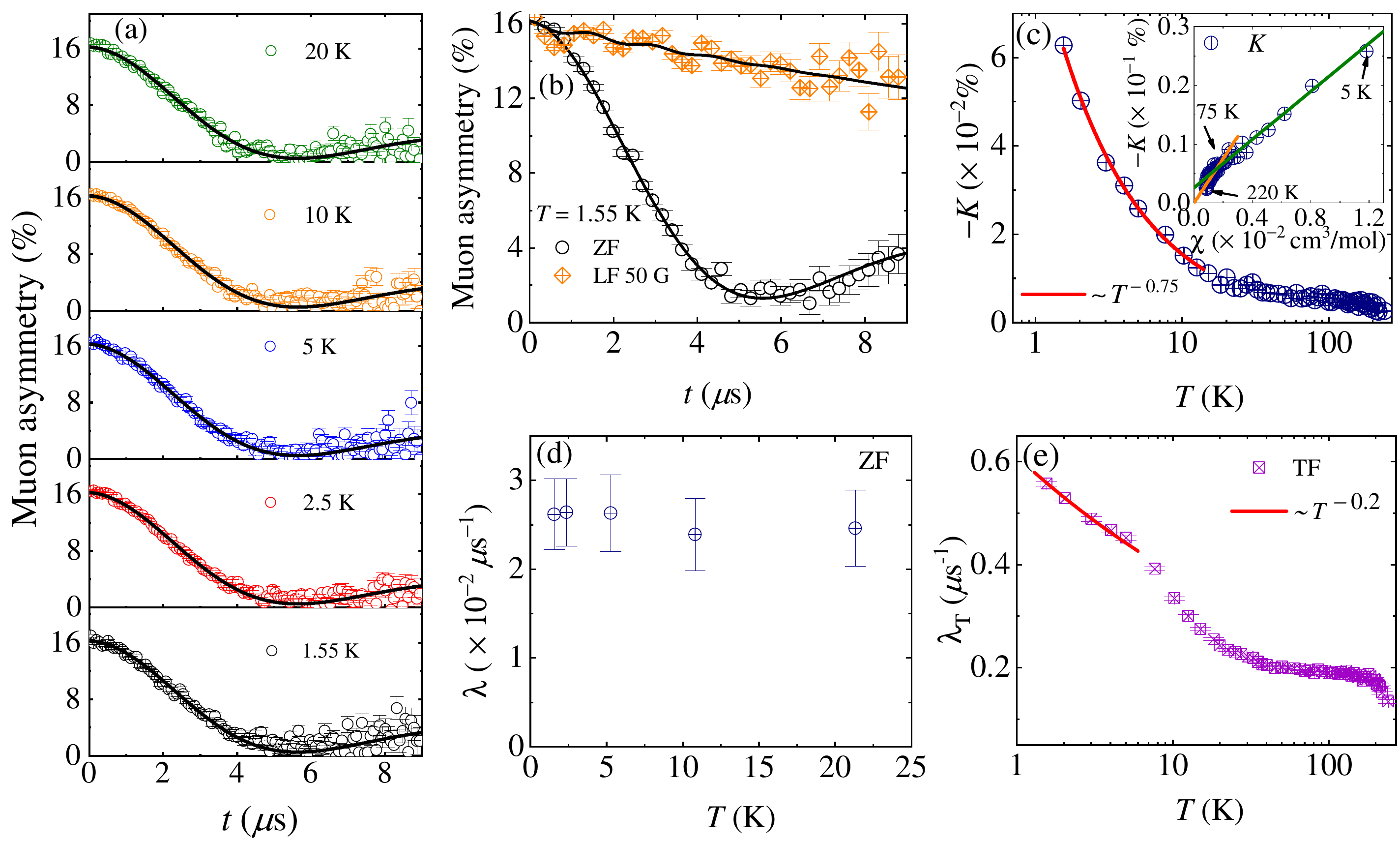}
	\caption{\textbf{Muon spin relaxation results  reveal  a  spin-liquid  state in Li$_{4}$CuTeO$_{6}$}.(a) The time evolution of zero-field muon spin asymmetry curves for several temperatures where solid lines are the fits to Eq. (\ref{eq:a1}).  (b) The time evolution of muon spin asymmetry in zero-field and in a longitudinal field (LF) of 50 G at 1.55 K with fits as discussed in the text. The simultaneous  fit of zero-field and  longitudinal field data to the Eq. (\ref{eq:a1}) yields $\sigma = $ 3.66 G and $\lambda$ = 0.029 $\mu$s$^{-1}$.  (c) The  temperature dependence of Knight shift where red line is  the fit of low-temperature  Knight shift i.e., $K \sim $ $T^{-\gamma}$  with $\gamma$ = 0.75  in the applied field of 4900 G. Inset shows the scaling of the muon Knight shift ($K$) with bulk susceptibility with temperature as an implicit parameter. The solid line is the linear fit in two regions (see text). (d) The temperature dependence of the longitudinal muon spin-relaxation rate in zero-field.  (e) The temperature dependence of the muon relaxation rate in the presence of a transverse field (TF) of 4900 G where red line is the power-law $T^{-0.2}$ fit in the low-temperature region. }{\label{Li$_{4}$CuTeO$_{6}$_muon}} 
\end{figure*}
In order to gain further insights into the magnetic properties relevant to this frustrated magnet and understand the ground-state degeneracy, it is important to extract the magnetic specific heat. Therefore, we have considered the zero-field specific heat data ($C_{\rm p}$) as a sum of intrinsic magnetic specific heat ($C_{\rm mag}$) due to exchange-coupled Cu$^{2+}$ ions, and lattice specific heat ($C_{\rm lattice}$) due to phonons. 
	After subtracting the lattice  contribution, the obtained magnetic specific heat exhibits a clear broad maximum (SI Fig.~6) around 0.4 K releasing $\sim 13{\text{\%}}$ of entropy [inset in Fig. 3(c)], which is a typical signature of short-range spin correlations in spin liquid candidates \cite{PhysRevLett.125.117206,Kitagawa2018}.
Comparing this with ED results on the high-$T$ model with no inter-chain coupling, which shows no such peak in zero field, and instead exhibits residual entropy at absolute zero due to unavoidable Kramers ground-state degeneracy of individual spin-chain fragments (see SI), this clearly shows the dominant impact of frustrated inter-chain couplings $J_\mathrm{inter}$ at low $T$.
	 Furthermore, it is observed that the temperature dependence of $C_{\rm mag }/T$  shows a $T^{-0.5}$ power-law behaviour (Fig.~\ref{Li$_{4}$CuTeO$_{6}$3} (c)) at low temperature, which suggest the presence of a random singlet state with strong low-lying excitations \cite{Kitagawa2018,PhysRevB.103.L241114}. Next, the entropy was obtained by integrating $C_{\rm mag}/T$ in the temperature range 0.052 K $\leq$ $T$ $\leq$ 2 K and is shown in  the inset of Fig.~\ref{Li$_{4}$CuTeO$_{6}$3} (c), with the contribution to entropy from specific heat below at $T$ below the lowest measured $T$ = 52 mK, estimated as $S  ({52} \ {\rm mK}) = C_\mathrm{mag}({52} \ {\rm mK})/(0.5)$. The saturation entropy amounts to just 13 \text{\%} of expected total entropy for a spin-1/2 system, which could suggest the presence of a highly degenerate ground state, but also naturally arises just from the fact that many spin degrees of freedom inside individual spin-chain fragments already freeze out at these low $T \ll J, J'$, reducing the total number of available spin degrees of freedom and making individual spin-chain fragments behave as single effective spins in a frustrated random 3D spin-lattice formed by weaker inter-chain interactions $J_\mathrm{inter}$. The presence of these frustrated interactions is also seen from the shift of specific heat to lower $T$ than predicted by theory if they were not present (see SI for a detailed comparison with ED calculations). 
	The power-law behaviour and the absence of any long range ordering down to 52 mK in zero field data suggest that the antiferromagnet Li$_{4}$CuTeO$_{6}$ is a quantum spin liquid candidate. 
 \\
Recently, Kimchi \textit{et al}. suggested that quenched disorder in  spin-liquid candidates 
can form a network of local moments in a random-singlet phase \cite{PhysRevB.22.1305,PhysRevLett.48.344}  and the low-temperature specific heat data  recorded in various magnetic fields should collapse to a universal curve characterized by a parameter $\gamma$, where  $\gamma$ is defined by the relation $C_{p}/T \propto \frac{1}{(\mu_{0}H)^{\gamma}} F(\frac{T}{\mu_{0}H})$ and $F$ is a scaling function \cite{Kimchi2018}.
We scaled the  magnetic specific heat data of LCTO accordingly (see Fig.~\ref{Li$_{4}$CuTeO$_{6}$3} (d)) and we indeed find  data collapse for $\gamma $ = 0.85, which could be attributed to the presence of random-singlets in spin-liquid ground-state. 
A similar data collapse is reported in a triangular-lattice Y$_{ 2}$CuTiO$_{6}$ with $\gamma$ = 0.70, where 50\text{\%} of Cu$^{2+}$-sites are diluted by Ti$^{4+}$ ions that yields quenched disorder \cite{PhysRevLett.125.117206}.
\subsection*{Muon spin relaxation}
In order to gain microscopic insights into the intrinsic susceptibility and the ground-state spin dynamics, we performed zero-field (ZF) and transverse-field (TF) muon spin relaxation ($\mu$SR) measurements  on polycrystalline samples. $\mu$SR is a unique local-probe technique with a broad time window to track  electron spin fluctuations. Therefore, ZF-$\mu$SR is one of the best tools to gain comprehensive microscopic information on magnetic ordering and spin dynamics of frustrated quantum materials.  
In Fig.~\ref{Li$_{4}$CuTeO$_{6}$_muon} (a), we present ZF-$\mu$SR spectra at a few representative temperatures showing that muon asymmetry remains more or less unchanged  for all temperatures, which confirms the absence of a phase transition down to 1.55 K \cite{PhysRevB.99.054412}. The time dependence of the muon asymmetry provides the information concerning  internal field distributions at the muon stopping site. 
In multi-domain or polycrystalline samples, static local-field of nuclear origin  $B_{\rm loc}$ is randomly oriented with respect to the initial muon spin direction.
The corresponding local-field distribution can be approximated by a Gaussian shape. In this case the muon relaxation function $P_{z}$($t$) (also known as the Gaussian Kubo-Toyabe function (KT) is of the form
\begin{equation}
	P_{z}(t)= \mathrm{KT}(t)=  \frac{1}{3}+\frac{2}{3}(1-\sigma^{2}t^{2})e^{-\frac{1}{2}\sigma^{2}t^{2}},     
\end{equation}
where $\sigma$ 
gives the local-field  distribution width.\\
In LCTO, the presence of competing magnetic fields  arising from electron spins provides dynamic muon spin relaxation in addition to relaxation due to quasi-static nuclear fields. Namely, a damping factor exp($-\lambda(T) t$) characteristic of dynamical fields has to be added to the KT function to account for the experimental results. Therefore, the zero-field asymmetry spectra  were fitted by the damped
Gaussian Kubo-Toyabe relaxation function 
\begin{equation}
	A(t)=A_{0}\mathrm{KT}(t)e^{-\lambda(\textit{T})t}, \label{eq:a1}   
\end{equation}
where $\lambda(T)$  is the dynamical muon spin relaxation rate due to electronic spins. The fitting of the observed ZF asymmetry data (Fig.~\ref{Li$_{4}$CuTeO$_{6}$_muon} (a)) by  Eq.~(\ref{eq:a1}) yields the nuclear field  distribution width of approximately 3.66 G  and the dynamical relaxation rate $\lambda$ = 0.023 $\mu$s$^{-1}$. In addition, the strong change of asymmetry spectra between zero-field and a small longitudinal field of 50 G as shown in Fig~\ref{Li$_{4}$CuTeO$_{6}$_muon} (b) also supports this model, as the applied field decouples the relaxation due to small static nuclear fields while it does not affect the relaxation due to larger but dynamical fields of electronic origin.  The obtained temperature dependence of the muon spin relaxation rate due to dynamical fields in  ZF  remains constant down to 1.55 K (Fig.~\ref{Li$_{4}$CuTeO$_{6}$_muon} (d))  without any signature of magnetic ordering, which suggests a dynamic ground-state. We note that the presence of a disordered static local
field at the muon site generally  leads to the ``1/3''  tail in  zero-field muon decay asymmetry of powder samples that is a hallmark of static magnetism.   However, zero-field muon asymmetries do not show  the ``1/3'' tail, which  confirms the absence of spin freezing \cite{yaouanc2011muon}.  In LCTO, the absence of any oscillation in ZF spectra (see Fig.~\ref{Li$_{4}$CuTeO$_{6}$_muon} (a)) also corroborates the absence of long-range magnetic ordering down to 1.55 K.\\
Next,  we carried out muon rotation measurements in the presence of large transverse field (TF) $B_{T}$ = 4900 G, which is much greater than the local fields.
The observed experimental TF spectra (not shown) were fitted using an oscillatory Gaussian decaying function 
\begin{equation}
	A(t)=A_{0} cos (2\pi \nu(\textit{T})t-\phi)e^{(-\lambda_{T}(\textit{T})t)^{2}/2}. \label{eq:a}
\end{equation} Here  $\phi \sim  \pi/3$ is the initial phase of the oscillation,  $\nu$(\textit{T}) is the frequency of muon precession originating from the local magnetic field and external transverse magnetic field, whereas $\lambda_{T}(T)$ is the transverse muon spin relaxation rate. The temperature dependence of $\lambda_{T}(\textit{T})$ is shown  in Fig.~\ref{Li$_{4}$CuTeO$_{6}$_muon} (e).  Upon decreasing the temperature, the relaxation rate increases and a rather strong change of $\lambda_{T}$ is observed  in the temperature range 1.55 K $<$ $T$ $<$ 10 K, which suggests the presence of a crossover region \cite{PhysRevLett.116.107203}. Above $T$ $>$ 20 K, $\lambda_{T}$ varies very slowly with temperature as would be expected for a paramagnetic region. In the presence of an external transverse field ($B_{T}$), the total magnetic field experienced by muon can be defined as the sum of $B_{T}$ and $B_{\rm loc}$. So, the mean precession frequency can be defined as $\nu(T)= \gamma_{\mu}(B_{T}+B_{\rm loc}(T))/2\pi$, where only the  local magnetic field $B_{\rm loc}(T)$  changes with temperature ($\gamma_{\mu}= 2\pi \times
135.53 $ MHz/T is the muon  gyromagnetic ratio). The muon Knight shift ($K$) that is directly proportional to magnetic susceptibility ($\chi$)  via  $K$ = ($\nu$($T$) $-$ $\nu_{\rm 0}$)/$\nu_{\rm 0}$ = $A \chi$, where $\nu_{\rm 0}$ = ($\gamma_{\mu}/2\pi) $$B_{T}$ is the reference frequency and $A$ is the coupling constant between  the muon magnetic
moment and the electron magnetic moments, thus captures  intrinsic susceptibility relevant to the spin-lattice.  Inset of Fig.~\ref{Li$_{4}$CuTeO$_{6}$_muon} (c) depicts the muon Knight shift as a function of bulk magnetic susceptibility ($\chi$) with temperature as an implicit parameter.  It is seen that $K$ as a function of $\chi$ exhibits two linear regions that are consistent with  the observed crossover  region in temperature dependence of $\lambda_{T}$. Below $T<$ 75 K, the linear fit (orange line) yields a coupling constant $A$ = $-$114 Oe/$\mu_{\rm B}$ whereas we find $A$ = $-$211 Oe/$\mu_{\rm B}$ from the linear fit (olive line) above  75 K. Furthermore, it is observed that the Knight shift data follows $K \sim $ $T^{-0.75}$ power-law  behavior at low $T$ (see Fig.~\ref{Li$_{4}$CuTeO$_{6}$_muon} (c)), which is consistent with bulk magnetic susceptibility and confirms that such susceptibility is intrinsic to the system. This behavior  further corroborates the presence of ``random-singlets" in LCTO.  $\mu$SR results do not indicate any static local fields of electronic origin down to 1.55 K. The measured muon asymmetry decay in zero field is, in fact, dominated by static nuclear fields, while relaxation due to dynamical fields of electronic origin remains hindered behind the nuclear relaxation  due to exchange narrowing down to 1.55 K. This reveals spin fluctuations are fast even at $T/\theta_{\rm CW} \approx 0.01$, which again  suggests a dynamic ground-state in LCTO. We note that the zero-field muon spin relaxation rate due to electronic fields (Fig.~\ref{Li$_{4}$CuTeO$_{6}$_muon} (d)) is much lower than the transverse muon spin relaxation rate (Fig.~\ref{Li$_{4}$CuTeO$_{6}$_muon} (e)) and its temperature dependence is different. As the former relaxation is only due to field fluctuations, while the latter is affected also by static local-field distributions induced by the applied transverse field, it is the induced local fields that are responsible for the increase of the transverse relaxation rate with decreasing temperature.
	This increase is characterized by a power-law $T^{-0.2}$ at the lowest temperatures (Fig.~\ref{Li$_{4}$CuTeO$_{6}$_muon} (e)).
\section*{Discussion}
Our investigation reveals the presence of  a strong antiferromagnetic exchange interactions between spins of Cu$^{2+}$ ions constituting a random frustrated spin-chain model with weaker inter-chain exchanges couplings forming an effective frustrated random 3D lattice of spin chain fragments at low $T$. Remarkably, the material doesn't show any signature of a phase transition or spin-glass freezing down to at least 45 mK, which suggests a dynamical liquid-like ground-state in this antiferromagnet.
The scaling behavior of magnetization and specific heat data at low $T$ reflects the presence of ``random-singlets" in a quantum disordered ground-state of LCTO. Magnetic materials with a dilute random network of $S$ = 1/2 sites can be considered as a combination of two sub-systems where the majority of constituent spins develop spin-liquid or valence bond crystal, and the minority forms random-singlets, and leaves out a random network of $S$ = 1/2 sites. For such a system, in the ground-state,  spins do not undergo any long-range antiferromagnetic  order.  The spins that are strongly coupled rather constitute singlet pairs and the remaining spins are almost free due to the development of effective antiferromagnetic exchange interaction ($J$) with a power-law probability distribution, $\mathcal[P](J) \sim J^{-\gamma}$, between them in the low-$T$ renormalization group flow of the random-singlet phase, where $\gamma$ is a positive quantity related to the spin correlation length \cite{PhysRevX.8.031028}. The small sub-system of spin-1/2 moments in the low-energy limit leads to unconventional scaling  behavior in  specific heat $C[T] \sim T^{1-\gamma} $ and spin susceptibility $\chi [T]$
$\sim T^{-\gamma}$ \cite{Kimchi2018}. 
In the presence of an applied magnetic field, specific heat captures the modified distribution of energy through a field-dependent coefficient, i.e., $C \sim T/(\mu_{0}H)^{\gamma}$ at low-temperatures.   It is worth to mention that the temperature and field dependence of magnetic  specific heat displays $T/\mu_{\rm 0}H$ data collapse for $\gamma$ = 0.85 in LCTO, which is a bit higher than the observed $\gamma$ = 0.5 for the honeycomb-lattice H$_{3}$LiIr$_{2}$O$_{6}$ and triangular-lattice LiZn$_{2}$Mo$_{3}$O$_{8}$ but it is close to the observed $\gamma$ = 0.7 for triangular-lattice Y$_{2}$CuTiO$_{6}$. 
The absence of strong spin-orbit interaction in Y$_{2}$CuTiO$_{6}$  and LCTO possibly  plays a vital role resulting in  data collapse with higher value of $\gamma$ compared to H$_{3}$LiIr$_{2}$O$_{6}$ and Li$_{2}$Zn$_{2}$Mo$_{3}$O$_{8}$. In addition, we also found a divergence in  the low-temperature
susceptibility of LCTO, which scales  as $T^{-\gamma}$ with $\gamma$  = 0.82  for bulk measurements and $\gamma$ = 0.75 for local-probe muon Knight-shift measurements. This value of  $\gamma$ is comparable with the reported  $\gamma$ = 0.68 and 0.76 for Y$_{2}$CuTiO$_{6}$ and Sr$_{2}$CuTe$_{1-x}$W$_{x}$O$_{6}$, respectively \cite{PhysRevB.102.054443,PhysRevLett.125.117206,PhysRevX.8.041040}. Further theoretical and experimental studies are desired to shed additional insights into the magnetism and spin dynamics of this promising spin-liquid candidate and to clarify the degree to which the random-singlet picture might be modified due to the presence of an underlying spin-chain fragment structure with random connectivity.
	The chain fragments could conceivably contribute further ingredients to low temperature physics of LCTO, leading to novel phenomena outside the paradigm of the usually considered random-singlet states. 
\vspace*{0.5cm}   
\section*{Conclusion}
We have investigated the structure, thermodynamic and local magnetic properties of  Li$_{4}$CuTeO$_{6}$ by employing X-ray, neutron diffraction, magnetization, specific heat, ESR, and muon spin
relaxation measurements. These were supported by state-of-the-art \textit{ab initio} density functional theory (DFT) and exact diagonalization (ED) calculations to elucidate the high-$T$ spin model and its thermodynamics. Li$_{4}$CuTeO$_{6}$ crystallizes in the monoclinic space group
$C$2/$m$, wherein the Cu$^{2+}$ site constitute spin chains with exchange $J$ along the [101] direction of random length due to partial occupancy of the Cu$_2$ site, with random additional Cu$^{2+}$ ions on partially occupied Cu$_3$ sites surrounding the main chains at positions displaced along $\pm b$ from the main-chain bond centers and coupling to two Cu$_2$ sites on the main chain through an exchange $J'$. The large and negative Curie-Weiss temperature, $\theta_{\rm CW}$ = $-$ 154 K,  suggests the  strong antiferromagnetic nature of the $J$ and $J'$ exchange interactions
between Cu$^{2+}$ moments, which is confirmed by DFT and fits of ED calculations to experimental data for $T >$ 20 K. These gradual deviation from these fits at low $T$ reveals further frustrating 3D interactions between the random spin-chain fragments, which dominate the low-$T$ response. Specific heat experiments reveal the absence of long-range magnetic ordering down to 52 mK despite the presence of  a large Curie-Weiss temperature, implying a 
highly frustrated spin-lattice. The absence of spin-freezing  despite  the  Li/Cu anti-site disorder reflects that Cu$^{2+}$ moments remain fluctuating down to 45 mK.  The zero-field and transverse-field $\mu$SR measurements
support a dynamic ground-state down to
1.55 K. 
Moreover, the existence of $T/\mu_{\rm 0}H$ scaling as well as magnetization and specific heat data collapse indicate the presence of a spin-1/2 network of random-singlets due to unavoidable anti-site disorder in the host lattice and frustrated weak 3D couplings. Our results demonstrate a randomness induced spin-liquid state in a frustrated magnet.
These results offer an exciting ground to explore novel  quantum states with unconventional low-energy excitations  in novel frustrated quantum materials with quenched disorder.
\section*{Methods} 
 Polycrystalline samples of Li$_{4}$CuTeO$_{6}$  were synthesized by the conventional solid-state reaction route as outlined in ref.~\cite{Kumar2012}. To obtain a single phase of LCTO, stoichiometric amounts of Li$_{2}$CO$_{3}$ (Alfa Aesar, 99.0 \text{\%}), CuO (Alfa Aesar, 99.995 \text{\%}), and TeO$_{2}$ (Alfa Aesar, 99.9995 \text{\%}) were mixed and the mixture was finely ground and pressed into a pellet, and after several intermediate steps, finally sintered at 850$^{\circ}$C for 30 h. The X-ray diffraction measurements were performed at 300 K by employing a Rigaku smart LAB X-ray diffractometer with Cu K$\alpha$ radiation ($\lambda $ = 1.54 {\AA}). Due to the light mass of Li and O atoms, the analysis of X-ray powder diffraction  data is not sufficient to determine the structural disorder  and the mixed site occupancy. Therefore, neutron diffraction measurements were performed at room temperature for incident wavelength of 1.28 {\AA} using crg-D1B Two-Axis Powder Diffractometer at  the Institut  of Laue-Langevin, France. Our neutron diffraction  studies detects neither spurious phases nor non-stoichiometry of oxygen. The neutron diffraction results at 1.28 {\AA} are shown in Fig.~\ref{Li$_{4}$CuTeO$_{6}$1} (c), while the neutron diffraction pattern, XRD pattern and resulting refinements parameters are presented in the Supplementary Information (SI). This suggests that  the polycrystalline samples used in this study are of very high quality. The refined  occupancy of all the atomic sites are within the experimental uncertainty of
about 1 \text{\%}, thus ruling out the presence of non-stoichiometric oxygen.\\  
Magnetization measurements were carried out using a Quantum Design, SQUID (MPMS)  in the temperature range 1.9 K $\leq \textit{T}\leq $ 300 K under magnetic fields up to 5 T. The electron spin resonance (ESR) spectrum was measured at 250 K at the irradiation frequency of 256.3 GHz on a custom built spectrometer with homodyne detection at the NHMFL, Tallahassee, USA. Thermoremanent measurements were performed using a Quantum Design, SQUID (MPMS) and the sample was cooled down from 100 to 2 K in either 1000 or 8 Oe, and turned down at 2 K. After 1h, the signal was measured again (at $t$ = 0). These results  rule out the presence of a spin-glass transition and are shown in SI Fig.~4.\\
 Specific heat measurements were performed in a Quantum Design, Physical Properties Measurement System (PPMS) by the thermal relaxation method, in the temperature range 1.9 K $\leq \textit{T}\leq $ 240 K and in magnetic  fields up to 7 T. Furthermore, specific heat  measurements were carried out separately in the temperature range 0.052 K $\leq$ $T$ $\leq$ 4 K in zero-field and 3 T  using a dilution refrigerator which was also used to measure ac susceptibility in the temperature range 0.045 K $\leq$ $T$ $\leq$ 4 K at four different frequencies  using a Dynacool PPMS instrument from Quantum Design, San Diego (USA).  \\ 
  $\mu$SR measurements were performed using the GPS spectrometer at the Paul Scherrer Institute, Villigen, Switzerland, on a polycrystalline sample with mass of approximately 1 g  down to 1.55 K  in zero field and in a transverse field  of 4900 G.\\
  DFT+$U$ calculations on a $2 \times 1 \times 2$ supercell of LCTO were carried out using the plane-wave DFT code~\cite{clark2005first} using a local (spin) density approximation (LSDA) exchange-correlation functional in a LSDA+$U$ scheme~\cite{anisimov1997first} with an effective Hubbard $U_\mathrm{eff}$ = 9 eV, a 2000 plane-wave energy cutoff, and a $2 \times 3 \times 2$ Monkhorst-Pack grid reciprocal-space sampling~\cite{monkhorst1976special}. To extract exchanges total energies of ${\geq}120$ random collinear spin configurations were calculated and fitted to a symmetry-adapted spin model~\cite{riedl2019ab}. See SI for further details. ED calculations with up to $N = 18$ spins were carried out on $>13,000$ distinct configurations of occupied Cu$_2$ and Cu$_3$ sites around finite-length spin-chain fragments in the high-$T$ random 1D model of Fig. 1(b), achieving a total probability between 60.2 \text{\%} and more than 98.8 \text{\%}, depending on parameters, that a randomly-chosen Cu$^{2+}$ spin in LCTO is part of one of the ED calculated spin-chain fragments. Details of these ED calculations are presented in the SI.
  
 \section*{Data availability}
 The data that support the findings of this study are available from the corresponding
 author upon reasonable request.  The neutron diffraction data can be found at  https://doi.org/10.5291/ILL-DATA.EASY-1043.
 
\section*{Acknowledgments}
We thank DST, India for the PPMS facility at IIT Madras. PK acknowledges the funding by the Science and
Engineering Research Board, and Department of Science and Technology, India through Research Grants.
AZ and MG acknowledge the financial support of the Slovenian Research Agency through Programme No. P1-0125 and Projects No. N1-0148 and J1-2461. A portion of this work was performed at the National High Magnetic Field Laboratory, which is supported by National Science Foundation Cooperative Agreement No. DMR-1644779 and the State of Florida. AMS thanks the NRF (93549) and the URC and FRC of the University of Johannesburg. Experiments at ILL was sponsored by the French Neutron Federation (2FDN), with data references from Easy Proposal N° 1043.
\section*{Author contributions}
All authors contributed substantially to this work.
\section*{Competing interests}
The authors declare no competing interests.
\section*{Additional information}
Correspondence and requests for materials should be addressed to P.K. (email: pkhuntia@iitm.ac.in)
\end{document}